\documentclass[showpacs,preprint,pre,floatfix,superscriptaddress]{revtex4}
\usepackage{ulem}
\usepackage{color}
\usepackage{graphicx}
\usepackage{amssymb}
\usepackage{amsmath}
\usepackage{amsfonts}
\usepackage{mathrsfs}
\renewcommand{\epsilon}{\varepsilon}

\begin{document}
\title{Polymer Translocation Induced by a Bad Solvent}
\author{Christopher L\"{o}rscher}
\affiliation{Department of Physics, University of Central Florida, Orlando,
Florida 32816-2385, USA}
\author{Tapio Ala-Nissila}
\affiliation{Department of Applied Physics, Aalto University School of Science 
and Technology,
P.O. Box 11000, FI-00076 Aalto, Espoo, Finland}
\affiliation{Department of Physics, Box 1843, Brown University, Providence,
Rhode Island 02912-1843, USA}
\author{Aniket Bhattacharya}
\altaffiliation[]{
Author to whom the correspondence should be addressed}
\email{aniket@physics.ucf.edu}
\affiliation{Department of Physics, University of Central Florida, Orlando,
Florida 32816-2385, USA}
\date{\today}
\begin{abstract}
We employ 3D Langevin Dynamics simulations to study the dynamics of polymer 
chains translocating through a nanopore in presence of asymmetric 
solvent conditions. Initially a large fraction ($>$ 50\%)  of the chain is placed 
at the \textit{cis} side in a good solvent while the $trans$ segments are placed in a 
bad solvent that causes the chain to collapse and promotes translocation from the $cis$ to 
the $trans$ side.  In particular, we study the ratcheting effect of a
globule formed at the \textit{trans} side created by the translocated segment,
and how this ratchet drives the system towards faster translocation.
Unlike in the case of unbiased or externally forced translocation
where the mean first passage time $\langle \tau \rangle $ is
often characterized by algebraic scaling as a function of the chain length $N$ 
with a single scaling exponent $\alpha$, and the histogram of the mean first 
passage time $P(\tau/\langle\tau \rangle)$ exhibits scaling, we find that 
scaling is not well obeyed. For relatively long chains we find
$\langle \tau \rangle \sim N^\alpha$ where $\alpha \approx 1$ for 
$\varepsilon/k_{B}T > 1$. In this limit, we also find that translocation proceeds with a nearly constant velocity 
of the individual beads(monomers), which is attributed to the coiling of the globule. We provide an approximate theory assuming rotational 
motion restricted on a
2D disc to demonstrate that there is a crossover from diffusive 
behavior of the center of mass
for short chains to a single file translocation for long chains, where
the average translocation time scales linearly with the chain length $N$.
\end{abstract}
\maketitle
\section{Introduction}
Attempts to understand the dynamics of viral invasion and infection
\cite{Alberts}, dynamics of DNA and other biopolymers passing through porous media 
(\textit{i.e.} cell
membranes), and direct medical applications such as gene therapy and
drug delivery, have rendered the field of polymer translocation a very
active field of research in recent years \cite{Meller_review,Muthu_review}.
Much work has been done to understand the physics involved in the translocation 
process. Analytical
work by Sung and Park \cite{Sung96}, Muthukumar \cite{Muthukumar00},
Chuang, Kantor and Kardar \cite{Chuang01,Kantor04},
Dubbeldam and coworkers\cite{Dubbeldam}, Panja and coworkers\cite{Panja}, and 
others\cite{Sakaue,Slater_1d,Milchev04}, 
supplemented by a vast amount of numerical work \cite{Luo1}-\cite{Bhattacharya_pre_2010},
has brought profound physical insight into the problem at hand.
In particular, much has been learned about making translocation faster in a 
controllable fashion, as this should be beneficial in biological systems.\par
To this end, a process known as
Brownian ratcheting \cite{Simon92} was discovered early on and 
discussed further in Refs.\cite{Peskin93}-\cite{Feynman}. Brownian ratchets are mechanisms by 
which translocation can be driven, or carried out more efficiently\cite{Feynman}. As
discussed by Simon, Peskin, and Oster\cite{Simon92}, thermal or chemical
asymmetries in the system can be used to extract useful work (\textit{i.e.} 
translocation of the polymer) from the thermal bath in accordance with the Second Law of
Thermodynamics \cite{Feynman}. As the polymer translocates, it experiences 
considerable back and forth motion due to thermal fluctuations. If the part
of the chain that is on the \textit{trans} side is modified in such a way as to
prevent backward motion through the pore, its random motion will be biased
and translocation through the pore is notably faster \cite{Gennes96}. This
modification to the chain which causes a biased translocation is often called a Brownian ratchet \cite{Simon92}.
A Brownian ratchet can manifest itself in many different ways. There can exist
binding particles that bind as chaperones on the \textit{trans} side
\cite{Zandi03,Bhattacharya_attr}, glycosylation can be used \cite{Gennes96}, 
or the chain can be tightly bound into
a coil on the \textit{trans} side via some method, usually by having
a bad solvent \cite{Wei07} or reducing the solvent's pH \cite{Gennes96}.
In a bad solvent, the polymer chain undergoes a
coil-globule transition to form a highly interacting spherical-like polymeric
configuration with a radius of gyration $R_g$ that scales with the number of 
monomers $N$ as $\langle R_g \rangle \sim N^{1/3}$\cite{Flory,Tanaka}.

In the present work, our aim is to study the influence of ratcheting on the 
dynamics of polymer
translocation as induced by an asymmetry in solvent quality between the \textit{cis} and \textit{trans} compartments. We focus on the 
case of two-sided translocation \cite{Luo1}, where a fraction of the polymer 
chain is initially placed at the $trans$ compartment, with good solvent on the \textit{cis} 
side and bad solvent on the \textit{trans} side. This asymmetry in the solvent condition 
induces a bias in the entropic barrier controlling translocation in such a way as to effectively drive the polymer to the \textit{trans} side. To characterize the dynamics of translocation, we analyze the waiting times, velocities, and 
effective forces on the individual monomers inside the pore. We find that for long chains with high attraction strengths, $\varepsilon/k_{B}T$, waiting times vary only slightly until the last beads emerge, at which point entropic effects become dominant. Since the velocity of the beads is inversely proportional to the waiting time, we get roughly constant velocity in this regime. This is  
apparent only for long chains with relatively high interaction strengths; for shorter chains with lower interaction strengths, the center of mass velocity of the polymer introduces an $N$ dependence into the velocity that cannot be 
overlooked when calculating how the average translocation time scales with 
$N$. We used this idea to make an approximate estimate for 
the $N$ dependence of the average translocation time. We find for short chains 
the translocation
exponent $\alpha \approx 2$, while in the large $N$ limit, we find $\alpha 
\rightarrow 1$.
We end our analysis by analyzing the histograms of the mean first passage 
time(MFPT). We note that since $\alpha$ 
varies between $2$ and $1$, we do not see universal scaling for a given value of 
$\varepsilon/k_{B}T$. However, in the large $N$ limit, we begin to see scaling 
manifest itself clearly, with a scaling exponent approaching the predicted value 
of unity.

\section{Model}

We use the Langevin equation to study the Brownian motion of particles in
solution. It is a statistical, stochastic differential equation of the form for 
each bead $i$:
\begin{equation}
m\ddot{{\bf r}_{i}}(t)=-\nabla U_{i}-\Gamma \dot{{\bf r}_{i}}(t)+ {\bf 
W}_{i}(t),
\end{equation}
where the total interaction,
\begin{equation} U_{i}=U_{FENE}^{i}+\sum U_{LJ}^{ij}\;,
\end{equation}
is the sum of the finitely extensible nonlinear elastic(FENE) spring potential interaction \cite{Grest86}
\begin{equation}U_{FENE}(r_{ij})=-\frac{1}{2}kR_{0}^{2}\ln\left[1-\left(\frac{r_{ij}}{R_{0}}\right)^{2}\right],
\end{equation}
and the
Lennard-Jones interaction between neighboring monomers,
\begin{equation}
U_{LJ}(r_{ij})=4\varepsilon_{ij}\left[\left(\frac{\sigma_{ij}}{r_{ij}}\right)^{12}-
\left(\frac{\sigma_{ij}}{r_{ij}}\right)^{6}-\left(\frac{\sigma_{ij}}{r_{ij}^{c}}\right)^{12}+
\left(\frac{\sigma_{ij}}{r_{ij}^{c}}\right)^{6}\right].
\end{equation}
The term ${\bf W}(t)$ describes the influence of Markovian white noise due to
the solvent, which is not taken into account explicitly here.
It satisfies the fluctuation-dissipation relation
\begin{equation}
\langle {\bf W}(t) \cdot {\bf W}(\tau) \rangle = 6k_{B}T\Gamma\delta_{ij}\delta(t-\tau).
\end{equation}
To model asymmetric solvent conditions on
the \textit{cis} and \textit{trans} sides, interaction cut-off values were
modified using a cut-off matrix.
Each particle was given a label (either 1 or 2, depending on whether it was on
the \textit{cis} side  or \textit{trans} side, respectively).
The cutoff values for the \textit{trans} side were set higher
($r_{c,22}=2.5\sigma$) than those for
the \textit{cis} side $(r_{c,11}=r_{c,12}=r_{c,21}=2^{1/6}\sigma)$. This ensures
that the monomers on the \textit{trans} side interact with a ``bad'' solvent\cite{Theta}, 
while the monomers on the \textit{cis} side are in a good solvent characterized by a self avoiding
random walk with Flory exponent $\nu=0.588$ ($R_g \sim N^\nu$) in 3D. 
The different cut-off values introduce a chemical potential difference, $\Delta\mu$, between 
the compartments. We will study the corresponding solvent asymmetry for
various monomer-monomer interaction coupling strengths, $\varepsilon$. In
particular, we will study how this solvent quality asymmetry drives the system towards a
much faster translocation.\par
The purely repulsive wall consists of one monolayer of immobile LJ particles of diameter 1.5$\sigma$
on a \textit{triangular lattice}
at the $xy$ plane at $z=0$. The pore is created by removing the particle at the center.
The reduced units of length, time and temperature are chosen to be  $\sigma$,
$\sigma\sqrt{\frac{m}{\epsilon}}$, and $\epsilon/k_B$ respectively.
For the spring potential we have chosen $k=30$ and $R_{ij}=1.5\sigma$, the friction coefficient
$\Gamma = 1.0$, and the temperature is kept at $1.5/k_B$ throughout the simulation. 

For a chosen fraction of the monomers at the $cis/trans$ 
we equilibrate the chain for a time on the order of the
Rouse relaxation time $\tau\sim N^{1+2\nu}$, where the Flory exponent $\nu=0.588$ in 3D . The chain is then allowed to
translocate using a time step of $dt=0.005$. As the last bead exits the pore,
a translocation event is completed and the process repeated for $2000$ times for 
averaging.

One way to understand the dynamics of a polymer chain translocating under such highly asymmetric conditions, is to study the analytic form of its free energy. Following Muthukumar, the 
free energy for $m$ translocated monomers is given by \cite{Muthukumar99}
\begin{equation}
\frac{\mathcal{F}_{m}}{k_{B}T}=(1-\gamma^\prime_{2})\ln(m)+(1-\gamma^\prime_{1})\ln(N-m)+m\frac{\Delta\mu}{k_{B}T}.
\end{equation}
Here, $\gamma^\prime$=0.5, 0.69, and 1 for Gaussian, self-avoiding, and rod-like chains, respectively. Driving force is easily obtained from the free energy by differentiating with respect to monomer index:
\begin{equation}
\frac{1}{k_{B}T}\frac{\partial{\mathcal{F}_{m}}}{\partial{m}}=(1-\gamma^\prime_{2})\frac{1}{m}+(1-\gamma^\prime_{1})\frac{1}{(m-N)}+\frac{\Delta\mu}{k_{B}T}.
\end{equation}
This is the driving force of the system in units of $k_{B}T$ for particular values of $\gamma^\prime_{1}$, $\gamma^\prime_{2}$, and $\Delta\mu$. As we shall demonstrate below, there is a delicate balance between the frictional and driving forces that will tend to set the system at a constant velocity.

The translocation time for a chain is a function of the number of monomers
on the \textit{trans} side at the beginning of the translocation
process $N_{tr}(t=0)$.
If we start our simulation having 50\% of the chain on the {\it trans} side, 
corresponding
to the ``two-sided" translocation first considered in Ref. \cite{Luo1}, 
\textit{i.e.}
%
%
${N_{tr}(t=0)}/{N} = 0.5$, this corresponds to releasing the chain down a downhill  
entropic barrier and therefore, the probability for successful 
translocation, $P(N_{tr}(t=0)$,
should be unity, which is indeed the case in our simulation. This probability 
decreases drastically
as the fraction $N_{tr}(t=0)/{N} $ is less than $0.5$, and especially
for long chains, the probability for a successful translocation is very small as 
shown in Fig.~1
for a chain of length $N=64$.
In the present work, we have studied the cases for
$N_{tr}(t=0)=0.5N$ and $N_{tr}(t=0)=0.25N$. By comparing these two sets of 
data,
we have extracted the results for the limit $N_{tr}(t=0) \rightarrow 0$. As 
expected, we
recover uniform scaling of the probability distribution for
the MFPT with a translocation scaling exponent $\alpha \rightarrow 1$ in the large $N$ limit.

\section{Results and Discussion}
To get an idea of the translocation process, we show typical snapshots of a 
translocating chain in Fig.~2(a) and Fig.~2(b) at different stages of the translocation 
process.
At $t=0$, the fraction of the chain that is located on the \textit{cis} side is 
characterized by the
equilibrium Flory exponent $\nu \simeq 0.588$, that corresponds to the good 
solvent condition.
Since the \textit{trans} part of the chain is in a poor solvent and the 
temperature is below the
$\Theta$-temperature \cite{Theta}, it will form a globule which is 
expected to grow as a function of time. Comparing the
snapshots for $\varepsilon/k_BT = 0.5$ (Fig. 2(a)) and 
$\varepsilon/k_BT = 1.5$ (Fig.~2(b)), we note that the
globule formed by the translocated segments becomes more compact as the
strength of the interaction increases.
We have checked the $N$ dependence of the radius of gyration
for chains immediately after the translocation process as shown in 
Fig.~3. For larger interaction strength, we find $R_g \sim N^{0.37}$, which is consistent 
with the
$N$ dependence of a compact spherical globule. It is worth mentioning that if 
all the globules for
different chain lengths
were perfect spheres and fully relaxed, then $R_g \sim N^{0.33}$. For $\varepsilon/k_BT = 
1.5$, we note from
the snapshots that the spheres are very compact and hence a dependence of $R_g \sim 
N^{0.37}$ implies
that the spheres formed by the translocated segments for different chain lengths 
are close to
equilibrium. For $\epsilon/k_BT = 0.5$, the corresponding exponent extracted from 
the slope of
$N=64,128$ and $256$ yields  $R_g \sim N^{0.26}$, which is less than $1/3$. This 
indicates that
the \textit{trans} side of the chain does not have sufficient time to relax during 
the
translocation process as discussed earlier in the 
literature\cite{Sakaue,Lehtola,Luo_EPL-09,Bhattacharya_pre_2010}.\par
We have monitored several quantities during
the translocation process.  First, we have analyzed the waiting time as a 
function of
monomer index (Fig.~4), where we
have defined waiting time to be the total time each bead spends at the pore
divided by the total translocation time $\langle \tau \rangle$ 
for the whole chain to cross the pore (\textit{i.e.}
$W(m)={\langle \tau(m) \rangle }/{\langle \tau \rangle}$), where $\tau(m)$
is the total time bead $m$ spends at the pore). The
notation $\langle \cdot \cdot \rangle$ indicates ensemble average over $2000$ 
iterations.
We notice that $W(m)$ initially increases and then decreases to almost zero, and the 
position of the
maximum increases with the chain length $N$. We also find that the peak position 
is an increasing
function of the interaction strength $\varepsilon/k_{B}T$.
Previously, waiting time of a monomer for a translocating chain was 
studied in great detail\cite{Luo1}. For a 
homopolymer undergoing
externally forced translocation, the residence time increases and becomes maximum 
for a monomer
index $m_{max} > N/2$; it then decreases (more rapidly than the rise) almost 
linearly, the position
of the maximum being skewed towards $m > N/2$.
When an attractive interaction is
present for the translocated segments, the barrier that 
the monomers at the \textit{cis} side are
pulled through is effectively skewed. However, this is different from applying a force only on 
the monomer inside
the pore. We note that data in Fig.~4 are similar to the case of forced 
translocation.
However, for the last $\sim$10\% of monomers, the residence time decreases very 
rapidly. This becomes more pronounced with increasing strength of the attractive 
interaction. It is evident from the residence time plots that longer chains with larger attraction 
strengths result in residence times that vary only slightly over the trajectory of the translocation, 
until the last beads emerge. We note that the waiting time of each monomer is inversely proportional 
to the velocity of that particular monomer at the pore. We have plotted the reciprocal of the waiting 
time function (scaled by the appropriate factors) for a chain length of 
$N=128$ (Fig.~5). We note that the velocity and inverse waiting time function collapse onto the same graph and are 
relatively constant up to the last few beads, at which point they both drastically increase due 
to entropic forces. We will explore the velocity at the pore for all the chains lengths below.

The behavior of the residence time when translated to the average velocity of 
the monomer
at the pore for various chain lengths $N=16 - 256$ and for
${\varepsilon}/{k_{B}T}=0.5$, $1.0$ and $1.5$ is shown in Figs.~6 
(a)-(c).
As predicted from the residence time
plots, the velocity for long chains is virtually constant for the entire translocation process
until the end, when the last beads emerge from the pore. We note that the driving force should be proportional to the velocity. From Eqn. (7) given for the driving force, we note that the term involving the reciprocal of the difference $N-m$, blows up as the last few beads emerge from the pore, explaining why velocity increases for the last few beads at the pore. This is most evident
for longer chains with higher $\varepsilon/k_{B}T$ (Fig.~6(b)-(c)).

To further understand our results, we have studied the force experienced at the
pore as a function of the monomer index Fig.~7(a)-(c). 
In order to get a better idea about the interaction of the chain with different solvents
on either side of the pore we have not shown the force arising out of the high frequency phonons 
from the anharmonic spring potential and have shown only the LJ contribution to the force in our plots.
For $\varepsilon/k_{B}T=0.5-1.5$ and relatively long chains, we see a rather flat force curve, which is close 
to
$F=0$, in agreement with $v=const$ discussed earlier (Fig.~7(a)). We interpret this result as being indicative of a force balance between frictional and driving forces 
$F_f$ and $F_{dr}$, respectively, at the pore. The driving force $F_{dr}$ is given by
$\frac{\partial \mathcal{F}_m}{\partial m}$ in Eqn. (7.)
%
%
For force balance to occur, we must then have the following condition:
\begin{equation}
\langle F_{dr}\rangle \approx \langle F_f \rangle .
\end{equation}
In the large $N$ limit, the driving force will be governed primarily by the chemical potential difference between the compartments $\Delta\mu$. In this limit

\begin{equation}
\langle F_{dr}(N \rightarrow \infty)\rangle \approx \Delta\mu \approx \Gamma 
\langle V(N \rightarrow \infty) \rangle .
\end{equation}
Thus, we see that the coiling velocity $v_c$ will be proportional to the chemical potential difference:
\begin{equation}
\langle v_c \rangle \approx \frac{\Delta\mu}{\Gamma} \approx const.
\end{equation} 

The main result is that in the large $N$ limit, the driving force and velocity are independent of $N$ and only dependent on the chemical potential difference between the compartments. This driving force is exactly balanced by the friction experienced by the monomers, which is proportional to the velocity of the beads. Thus we see that in the large $N$ limit the condition $\sum F \approx 0$ and 
$v \approx const$ is well obeyed.

In the case of short chains, the force becomes negative gradually for the last beads, 
whereas for long chains the change is rather drastic (Fig.~7(a)-(c)). We interpret this negative 
force as being a result of the last few beads, 
(having a relatively large velocity) while still being at the $cis$ compartment but 
in the vicinity of the pore, getting absorbed by the globule on the \textit{trans} compartment, resulting in a deceleration as they escape the $cis$ compartment. This results in a negative force
on the last few monomers.

Next we present a simple scaling argument for estimating
how the average translocation time should scale as a function of $N$. It is 
based on
the observation that the translocation dynamics
corresponds to a ``coiling" of the chain around the collapsed globule on the 
\textit{trans}
side with coiling velocity $v_c$ (Fig.~8). If we assume that the attractive force 
$F_e$ on the bad solvent side is directed towards the center
of the globule, we can write $F_{e} \sim m {v_c^{2}}/{R}$, which gives
$v_c \sim \left({RF_{e}}/{m}\right)^{{1}/{2}}$. For the collapsed globule close
to equilibrium $R \sim N^{1/3}$ and $m\sim N$, and thus
$v_c \sim\left(N^{-{2}/{3}}F_{e}\right)^{{1}/{2}}$. 
Due to spherical symmetry, we expect that the force will be proportional to the 
number of monomers in a disk of radius $R\sim N^{1/3}$. Thus, $F_{e}\sim 
N^{2/3}$, from which we extract that $v_c = const$. Now, $\langle \tau \rangle 
v \sim R_{g}^{3}$, where $R_{g}^{3}\sim N$ and $v ={1}/{N}+ v_{c}={1}/{N}+const.$ 
The ${1}/{N}$ dependence comes from the velocity of the center of mass, and will 
only contribute in the low $N$ limit. Thus we can see that in the low $N$ limit, 
the ${1}/{N}$ term dominates and thus $\langle \tau \rangle \sim N^{2}$, which 
is supported by our results below. On the other hand, for large $N$, the 
constant will dominate, thus arriving at the large $N$ limit of $\langle \tau 
\rangle \sim N$, also supported below.  Thus for single file translocation 
induced by coiling, we would expect a scaling exponent close to $\alpha \approx 1$, much
lower than 3D forced translocation where 
$\alpha \approx 1.37-1.6$ depending on the rate of translocation 
\cite{Luo_PRE-RC,Bhattacharya_epje09,Luo_EPL-09,Bhattacharya_pre_2010}.

We have tested the scaling argument above for several values of $\varepsilon$
by studying the histograms of the MFPT for several values of chain length, 
namely $N=$16, 32, 64, 128, and 256 to see how they scale as a function of $N$. 
Following previous work\cite{Bhattacharya_epje09} we have used nonlinear regressions 
of the form $f(x)=Ax^{B}\exp(-Cx)$. The maxima for these curves occur at $x={B}/{C}$. 
We have used the position of the maxima for each chain length to obtain the mean first 
passage time $\langle \tau \rangle $ from which we can extract the scaling exponents
given in Table I. Scaled MFPT histograms are shown in Fig.~9(a)-(c) for
$\varepsilon/k_{b}T=0.5-1.5$. We notice from Table I that the scaling exponent
$\alpha$ decreases when extracted from successive larger values of $N$. 
If we look at the spectrum of exponents calculated for short and long chains and for 
various values of the interaction strength $\epsilon$ qualitatively we notice that 
for short chains and weaker $\epsilon$ the exponent reflects diffusive behavior, while for 
larger combination of $N$ and $\epsilon$ the exponent is less than the corresponding exponent 
for forced translocation for similar chain lengths
\cite{Luo_PRE-RC,Bhattacharya_epje09,Luo_EPL-09,Bhattacharya_pre_2010}. In the next section 
we provide theoretical argument why the translocation behavior from short $N$ and low
$\epsilon$ is dominated by diffusive behavior while for long $N$ and large $\epsilon$, this 
diffusive behavior crosses over to a ``single file'' translocation asymptotically reaching 
a scaling exponent $\alpha \rightarrow 1$ in this limit. Evidently, for this reason we do not 
see data collapse of the scaled histogram for the MFPT accros the board. However, it is worth 
noticing that for long $N$ and large $\epsilon$ ($N=128$ and 256 and $\epsilon/k_BT = 1.5$) we notice 
a near perfect data collapse. One can see this trend from Eqn. (7).
For short chains we see that chain length plays quite an important role in determining 
the driving force on the polymer. For relatively long chains, as we showed earlier, 
this $N$ dependence is washed away, and the only contribution to the driving 
force is the chemical potential difference. Thus for $N \rightarrow \infty$ the 
driving force becomes independent of chain length. \par

We have also compared how the MFPT data collapse on a single
master curve when we use the initial condition,
$N_{t}(t=0)=N/4$ instead of $N_{t}(t=0)=N/2$ as shown in Fig.~10.
We notice that data collapse and scaling is more closely obeyed in this regime.
The scaling exponent for $\varepsilon/k_{B}T=1.5$ continues to suggest a 
tendency
towards $\alpha \rightarrow 1.0$ in the large $N$ limit, as is also predicted by Wei 
\textit{et al.}
in their studies of the effect of solvent quality asymmetries on the translocation process 
\cite{Wei07}, although studied differently and using a different model for the solvent conditions.
Their studies indicate that polymers translocating under different solvent 
qualities have a
scaling law that varies from $\langle \tau \rangle \sim N^{1+2\nu}$ to 
$\langle \tau \rangle \sim N$,
which is quite close to our results. Our present results are also consistent 
with Muthukumar's analytical expression for the translocation time as a function of $N$ 
for various conditions for the chemical potential difference \cite{Muthukumar99}. 
Calculations in Muthukumar's work show that for symmetric barriers, the translocation time 
will scale as $\langle \tau \rangle\sim N^2$. For asymmetric barriers and long chains, if the entropic terms 
in the free energy equation are small compared to the term involving $\Delta\mu$, 
then the translocation time scales linearly as $\langle \tau \rangle\sim N$ for $N\|\Delta\mu\|>1$, 
and scales as $\langle \tau \rangle\sim N^2$ for $N\|\Delta\mu\|<1$, which is consistent with our 
present results. For relatively short chains (\textit{i.e.} $N=16$) and weak coupling 
strength $\varepsilon$, we note a scaling exponent close to $\alpha \approx 2$. 
For longer chains with stronger coupling constants $\varepsilon$, we notice a trend 
towards unity, in complete agreement with Muthukumar's analytic expression.
We expect that for longer chains,
this limit will be even more closely reached due to the single file nature
of the translocation induced by the coiling of the globule on the bad solvent
compartment. This is also reflected in in Fig.~11(a)-(c) we show 
the translocation time $\langle \tau \rangle$ plotted against $N$ on a log-log scale. 
In each attempt to scale the MFPT histograms, we have used the linear fit slope of the 
entire log-log plot (\textit{i.e} $\langle \tau \rangle \sim N^{\alpha}$). The exponents support our 
claim that the the translocation interpolates from diffusive to a single 
file behavior.  
\section{Summary and Conclusions}
We have investigated using 3D Langevin Dynamics simulations the
properties of a homopolymer translocating through a nano-pore with a solvent 
asymmetry.
In our model, there is good solvent on the \textit{cis} side of the pore, while
on the \textit{trans} side there is bad solvent. This creates an effective driving 
force
on the polymer and leads to the emergence of a collapsed globule on the 
\textit{trans}
side during the translocation process. We have used a free energy argument 
to show that the driving force is relatively insensitive to chain length in the 
large $N$ limit 
and is governed mainly by the chemical potential difference between the compartments. 
As is evident from our force plots, there is a delicate balance between this constant 
(in the large $N$ limit) driving force and the friction force experienced by the beads. 
Consistent with this idea, we find that the velocity of the beads at the pore is 
relatively constant in the large $N$ limit, which is attributed the constant coiling 
velocity occurring on the "trans" compartment. Furthermore, we note that scaling is not 
well obeyed in the low $N$ limit as is evident from our mean first passage time 
histograms. 
We interpret this as being a consequence of the $N$ dependence of the driving 
force (and thus velocity!) in the low $N$ limit. For longer chains, however, 
we note that the driving force(and velocity) become quite insensitive to changes 
in the chain length, and we retrieve scaling with a scaling law close to our predicted 
large $N$ limit of $\langle \tau \rangle\sim N$. This can be interpreted as a crossover phenomenon 
from diffusive type translocation to a single-file driven translocation. Our 
studies might be relevant for translocation of biopolymers accros 
cell membranes. 
\section{Acknowledgements:} 
CJL at the University of Central Florida has been supported by the 
Florida Education Fund (McKnight Doctoral Fellowship). 
AB acknowledges funds from 
Department of Applied Physics, Aalto University School of Science 
and Technology 
for a visiting Professorship during summer 2009. TAN acknowledges travel 
fund from University 
of Central Florida during May 2009 and funds by the Academy of Finland through
the COMP CoE and TransPoly consortium grants.

\newpage
REFERENCES: \\

\newpage
\centerline{FIGURE CAPTIONS}

\noindent
Fig.~1:~Probability of successful translocation as a function of the initial 
number of monomers on the $trans$ side for a chain of length $N=64$.\\

\noindent
Fig.~2(a):~Snapshots of a translocating chain of length $N=256$ for 
$\epsilon/k_BT = 0.5$ at times (a) $t=0$, (b) $0.25\tau$, (c) $0.5\tau$, 
(d) $0.75\tau$, and (e) $1.0\tau$ respectively.\\

\noindent
Fig.~2(b):~Snapshots of a translocating chain of length $N=256$ for 
$\epsilon/k_BT = 1.5$ at times (a) $t=0$, (b) $0.25\tau$, (c) $0.5\tau$, 
(d) $0.75\tau$, and (e) $1.0\tau$ respectively.\\

\noindent
Fig.~3:~Variation of $\langle R_ g \rangle $ as a function of number of translocated 
monomer $N_{tr}$ (log-log plot) for $\varepsilon/k_BT = 1.5$ (black circles) and 
$\varepsilon/k_BT = 0.5$ (red squares) respectively. \\

\noindent
Fig.~4:~Average waiting time as a function of monomer index normalized by the
maximum waiting time for various chain lengths. The symbols circles (black), 
squares (red), diamonds (green), triangle up (blue), and triangle left (magenta) 
correspond to the chain lengths $N=16$, 32, 64, 128, and 256 respectively 
(color online) for (a) $\varepsilon/k_{B}T=0.5$.
(b) $\varepsilon/k_{B}T=1.0$, and (c) $\varepsilon/k_{B}T=1.5$.\\

\noindent
Fig.~5:~Inverse of waiting time (red diamonds) and velocity of the monomer beads 
(blue circles) plotted as a function 
of monomer index for chain length $N=128$ and $\varepsilon/k_{B}T=0.5$. 
We note that the two graphs almost fall on top of each other and are 
relatively constant until 
the end of the translocation, when the last beads emerge out of the pore. \\

\noindent
Fig.~6:~Average velocity on the monomer beads inside the pore as a 
function of monomer
index for chain lengths $N=16$, 32, 64, 128, and 256 respectively for 
(a) $\varepsilon/k_{B}T=0.5$, (b) $\varepsilon/k_{B}T=1.0$, and (c) $\varepsilon/k_{B}T=1.5$. 
The symbols have the same meaning as in Fig. 4. We note that in the large $N$ limit, 
the velocity of the monomers becomes almost constant excepting for the last few monomers. \\

\noindent
Fig.~7:~Average force on the monomer beads inside the pore as a function of monomer
index for chain lengths $N=16$, 32, 64, 128, and 256 respectively for 
(a) $\varepsilon/k_{B}T=0.5$, (b) $\varepsilon/k_{B}T=1.0$, and 
(c) $\varepsilon/k_{B}T=1.5$.
The symbols have the same meaning as in Fig. 4. 
To analyze the force, we have taken only the LJ
contribution (neglected the smallest the fluctuations). In agreement with
the velocity plots, we notice that the force at the
pore is not only constant, but also zero in the large $N$ limit, which agrees with the
constant velocity that we find at the pore, indicating a force balance at the pore 
between friction and driving force. We also notice that the force drastically becomes 
negative as the last few beads translocates. \\

\noindent
Fig.~8: Schematic of a polymer coiling ideally around a collapsed globule. 
The force is directed towards the center and is proportional the 
number of monomers in a 2D disc around which the translocating polymer coil.\\

\noindent
Fig.~9:~Scaled histograms for the MFPT for different chain lengths $N$=16, 32, 64, 128, 
and 256 for (a) $\varepsilon/k_{B}T=0.5$, (b) $\varepsilon/k_{B}T=1.0$, and 
(c) $\varepsilon/k_{B}T=1.5$ for $N_{tr}(t=0)/N=0.5$. The symbols have the same 
meaning as in Fig. 4. Here, we have used the linear fit slope for the entire Log-Log 
plots (Fig.~11) for $\varepsilon/k_{B}T=0.5-1.5$ as scaling exponents. 
In general for these values of $\varepsilon/k_{B}T$ we do not notice universal 
scaling for all chain length $N$. However, for longer chains and larger attraction strengths 
(inset in (b) and (c)), we begin to see scaling emerges 
quite clearly for $\alpha \approx 1$. \\

\noindent
Fig.~10:~(a) Unscaled and scaled MFPT Histograms for 
chain length $N=64$ (green diamonds) and $N=128$ (blue triangle-ups) with 
$N_{tr}/N(t=0)=0.25$ as the  initial condition. (a) for $\varepsilon/k_{B}T=0.5$  
where we notice a scaling exponent close to the value attained for the corresponding 
case of $N_{tr}/N(t=0)=0.50$ shown in Fig. 9. (b) For $\varepsilon/k_{B}T=1.5$;   
we notice a scaling exponent closer to the 
theorized large $N$ limit of unity. We also note that scaling is 
more closely obeyed in this case.\\

\noindent
Fig.~11:~Variation of $\langle \tau \rangle$ as a function of $N$ 
(log-log plot) for (a) $\varepsilon/k_{B}T=0.5$, 
(b) $\varepsilon/k_{B}T=1.0$, and (c) $\varepsilon/k_{B}T=1.5$ respectively.
The local values of the slope ($\alpha$) are indicated in the graph.

\newpage
\begin{table}[ht!]
\begin{center}
Table I:~Effective scaling exponents ($\langle \tau \rangle \sim N^{\alpha})$ 
for various combinations of $N$ and $\frac{\varepsilon}{k_{B}T}$ \\ 
\vskip 0.5truecm
\begin{tabular}{|c|c|c|c|}
\hline
$N$ & $\frac{\varepsilon}{k_{B}T}=0.5$ & $\frac{\varepsilon}{k_{B}T}=1.0$ &
$\frac{\varepsilon}{k_{B}T}=1.5$\tabularnewline
\hline
\hline
16-32 & 2.0 & 1.5 & 1.3\tabularnewline
\hline
32-64 & 1.6 & 1.10 & 1.1\tabularnewline
\hline
64-128 & 1.3 & 1.10 & 1.2\tabularnewline
\hline
128-256 & 1.2 & 1.1 & 1.2\tabularnewline
\hline
\end{tabular}
\end{center}
\end{table}
\vskip 5.0truecm
\vfill
\newpage
\begin{figure}[ht!]
\begin{center}
\vskip 1.0truecm
\includegraphics[width=13.0truecm]{Fig1.eps}
\vskip -4.0truecm
\label{prob}
\end{center}
\vskip 5.0truecm
\centerline{\large Fig. 1}
\end{figure}
\begin{figure}[ht!]
\begin{center}
\includegraphics[width=5.0truecm]{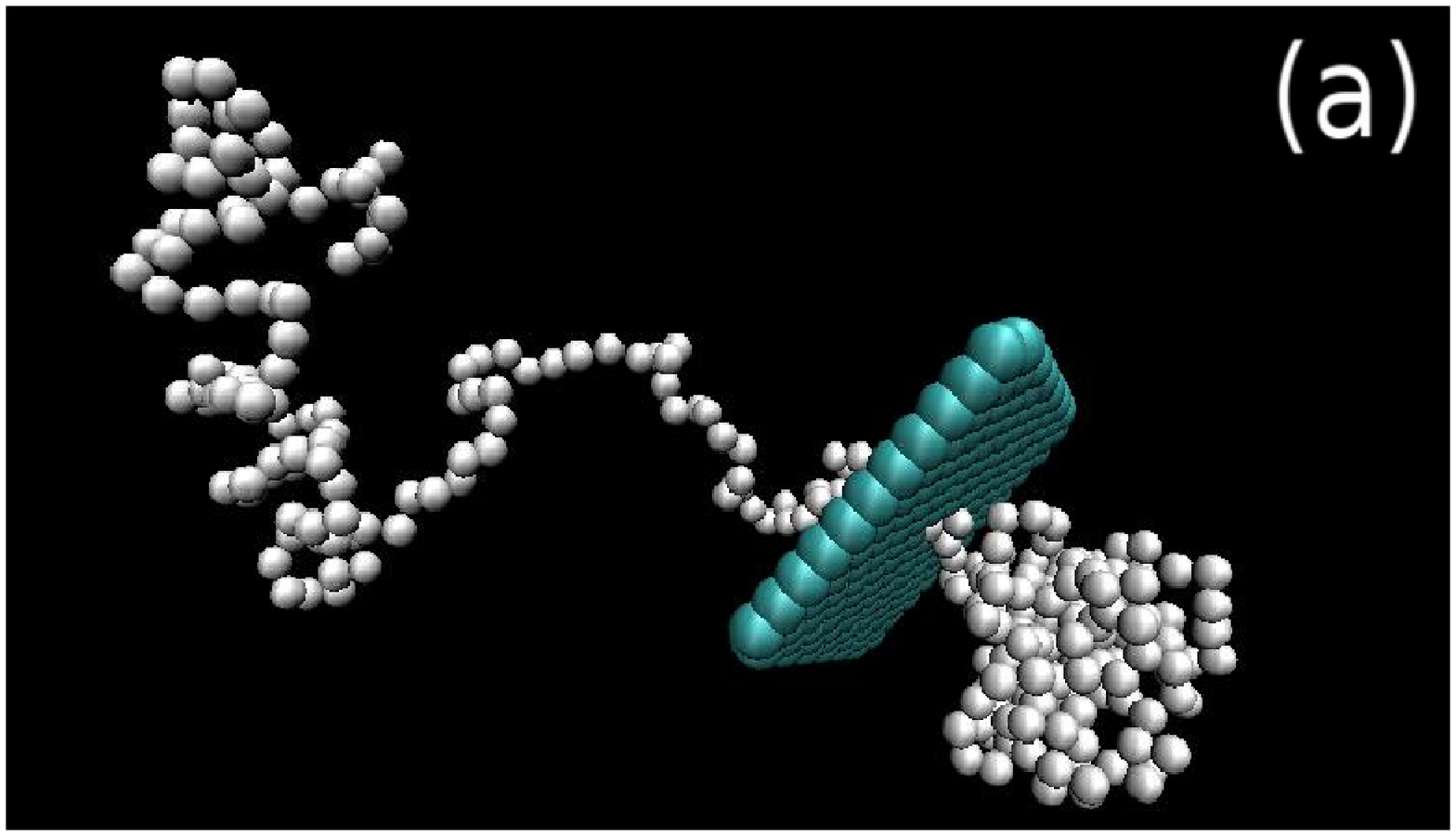}\\
\includegraphics[width=5.0truecm]{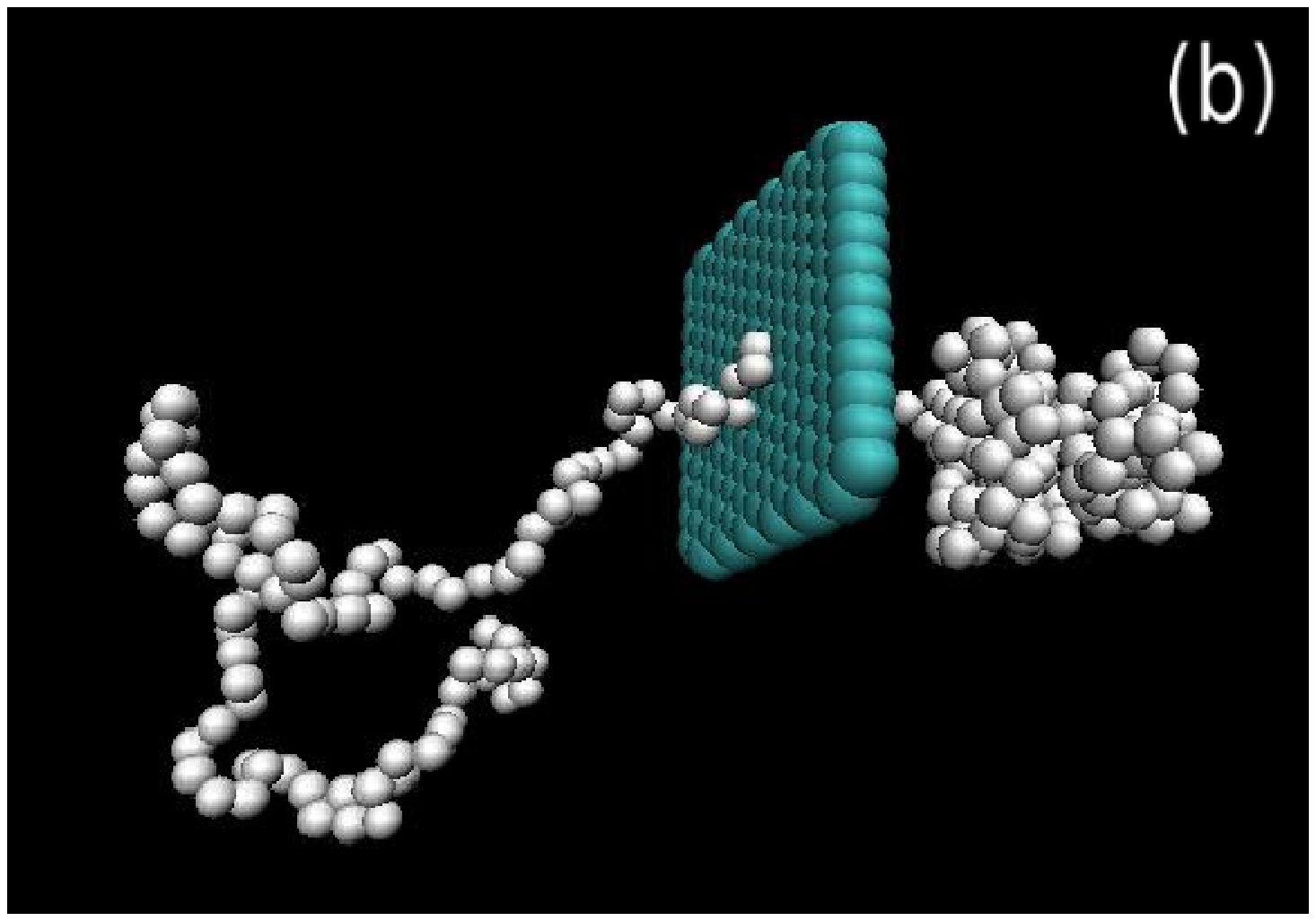}\\
\includegraphics[width=5.0truecm]{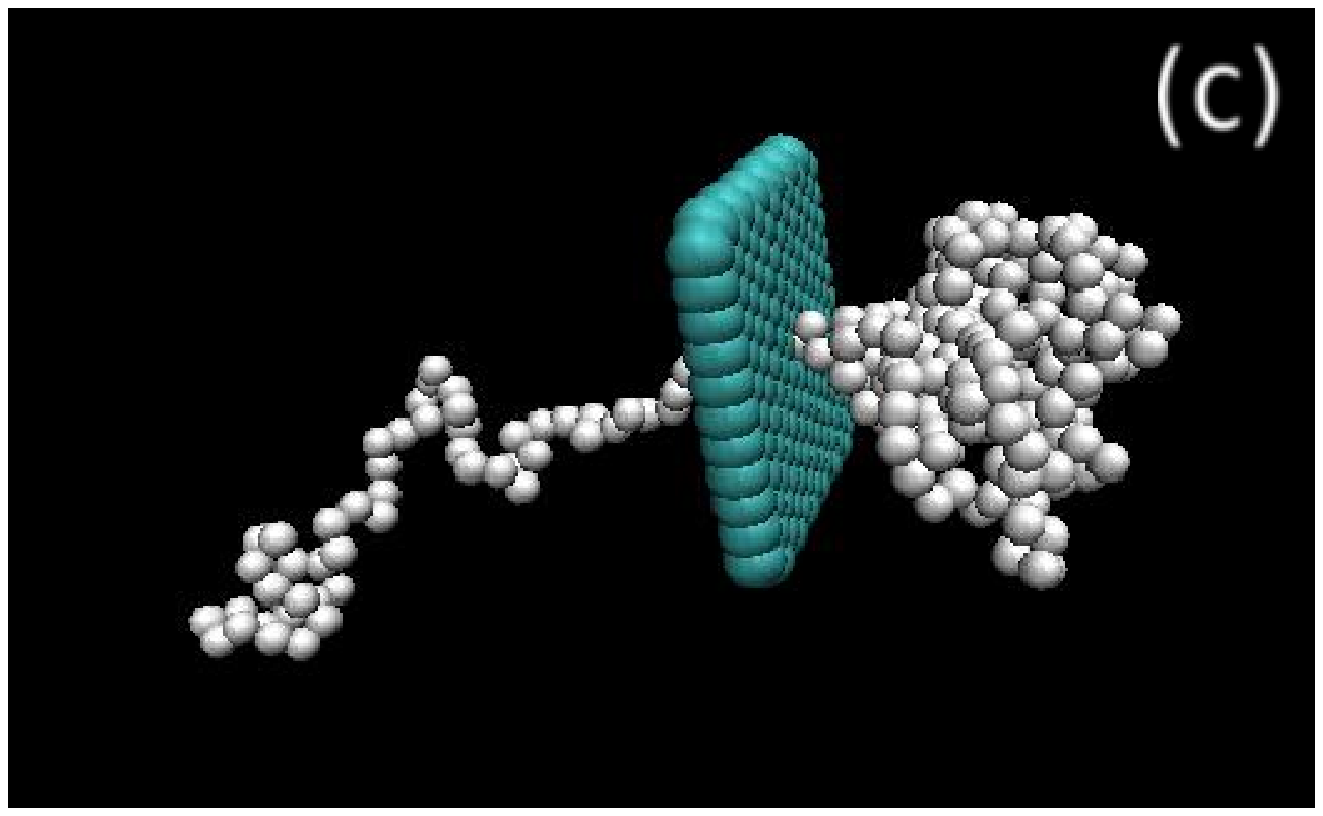}\\
\vskip 0.05truecm
\includegraphics[width=5.0truecm]{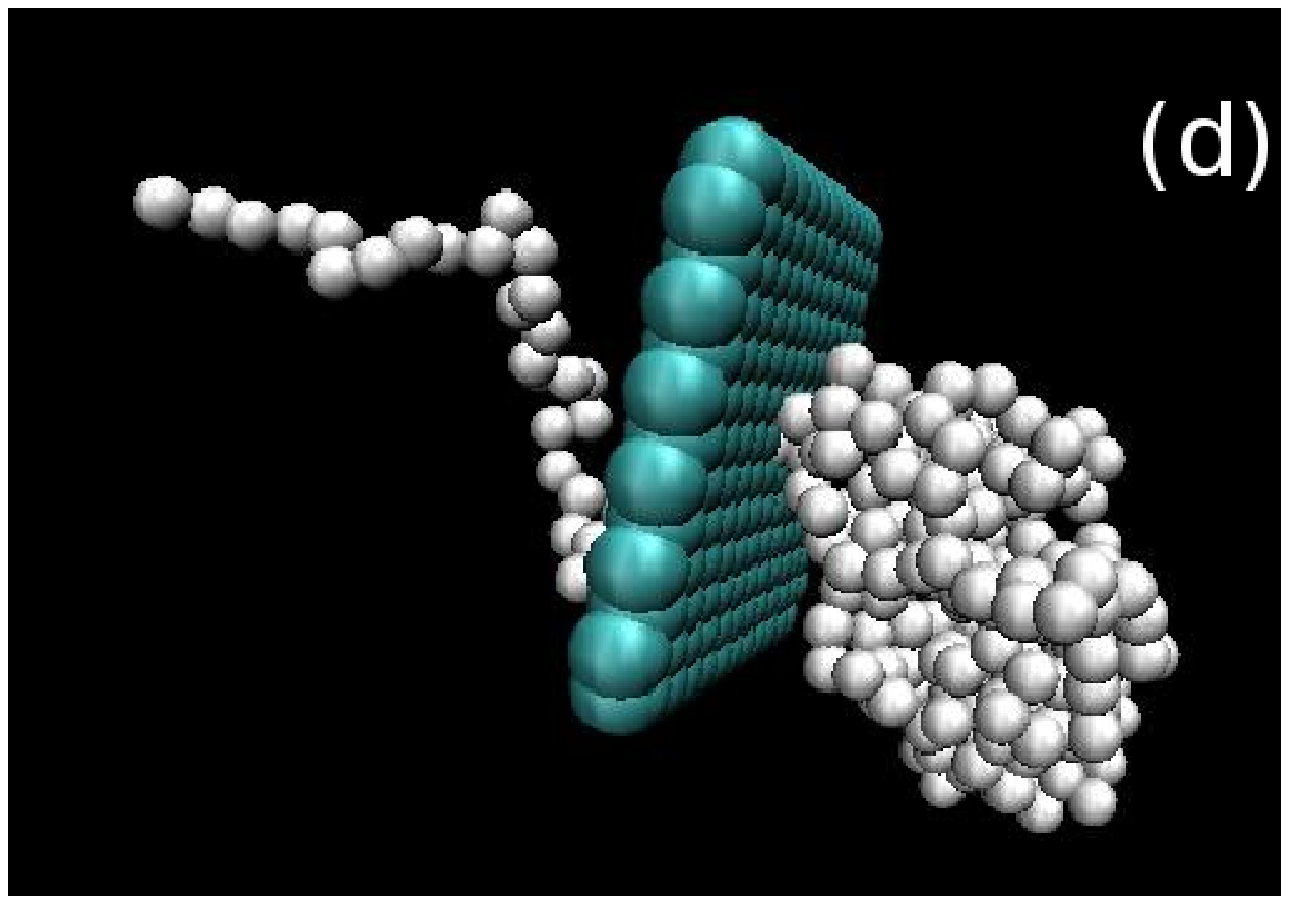}\\
\includegraphics[width=5.0truecm]{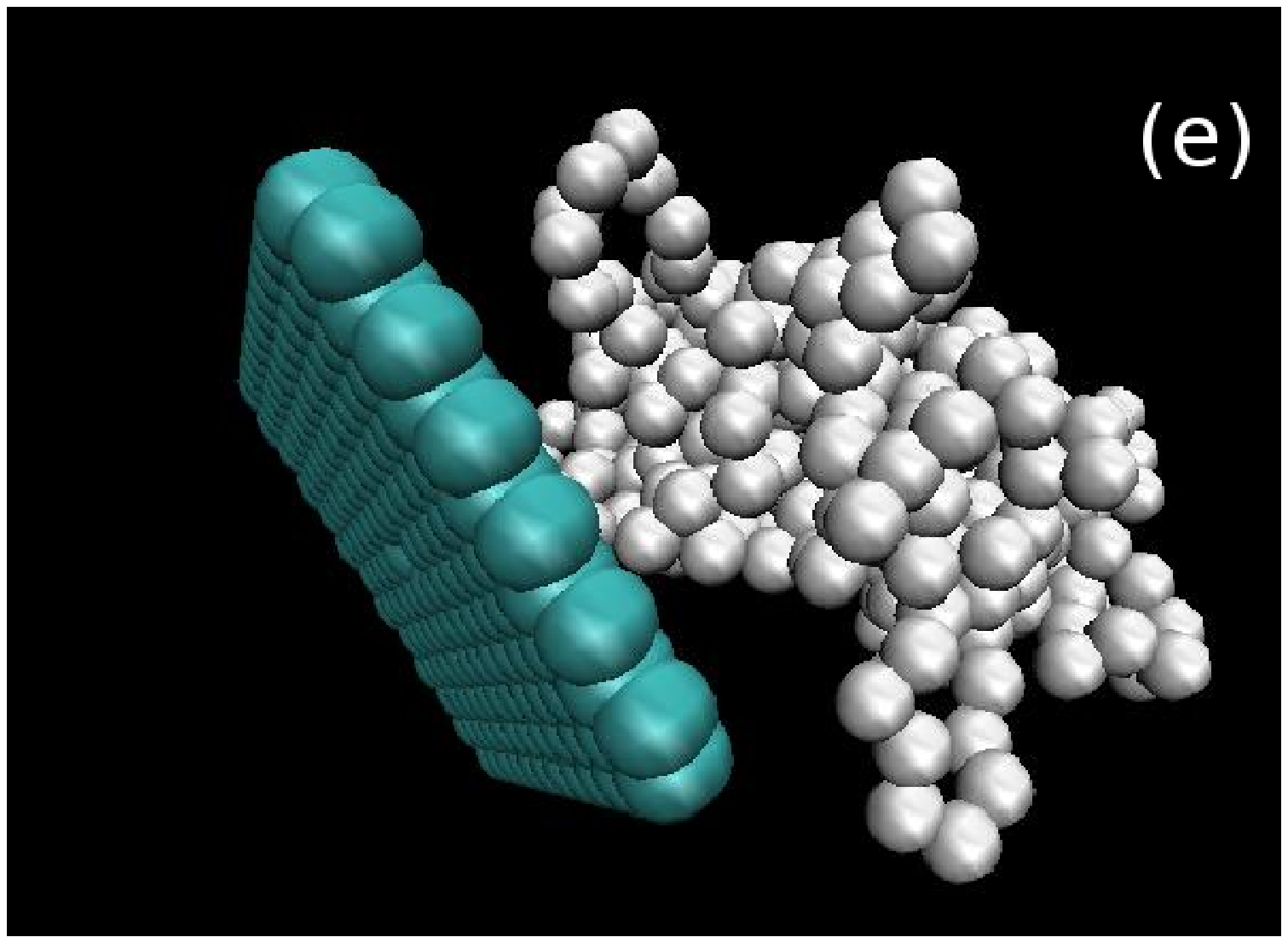}
\end{center}
\vskip 0.25truecm
\centerline{\large Fig. 2(a)}
\end{figure}
\begin{figure}[ht!]
\begin{center}
\includegraphics[width=4.0truecm]{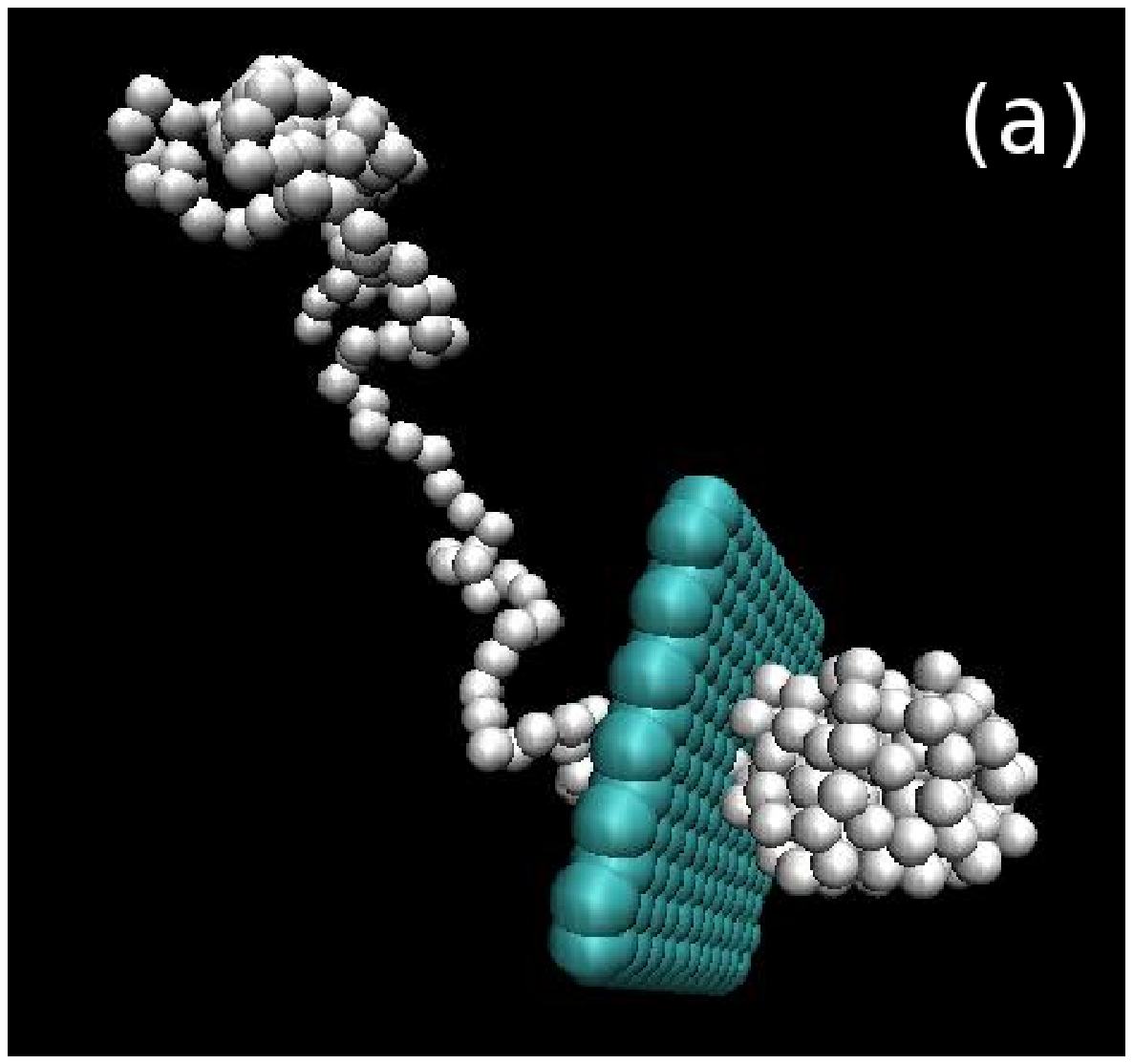}\\
\includegraphics[width=4.0truecm]{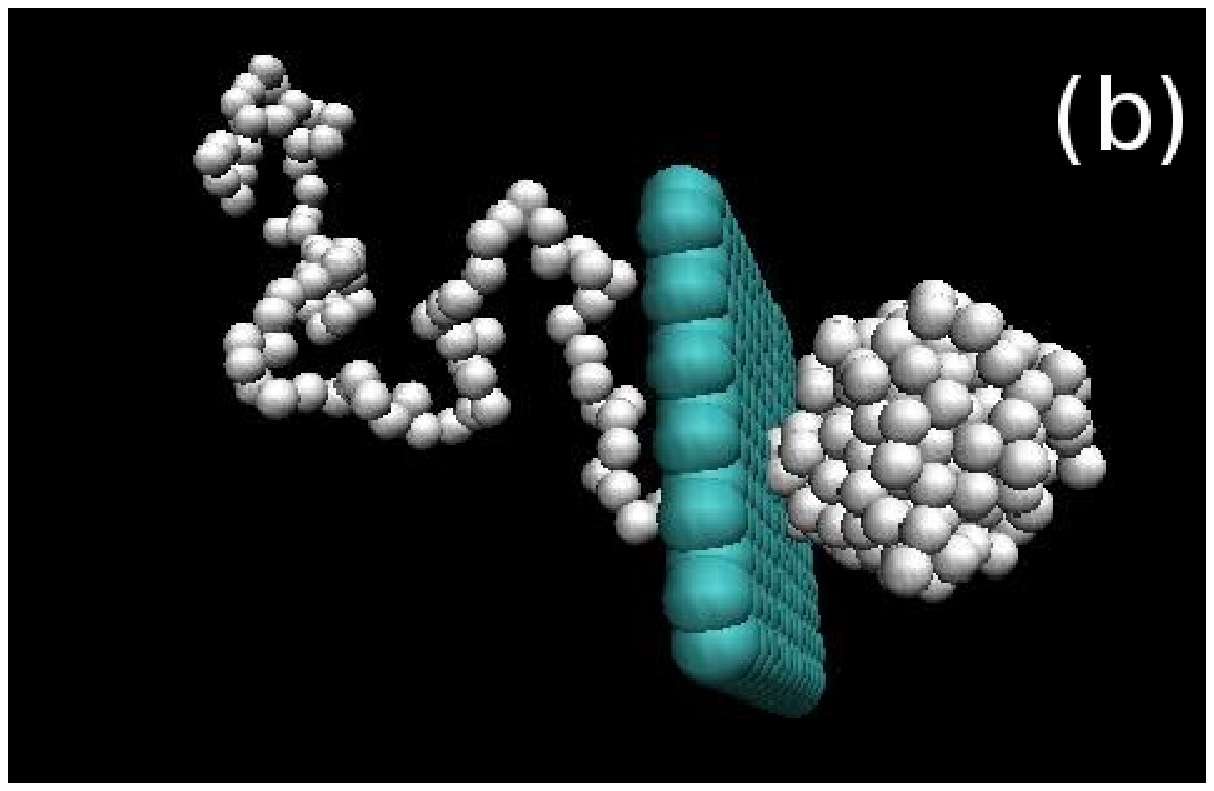}\\
\includegraphics[width=4.0truecm]{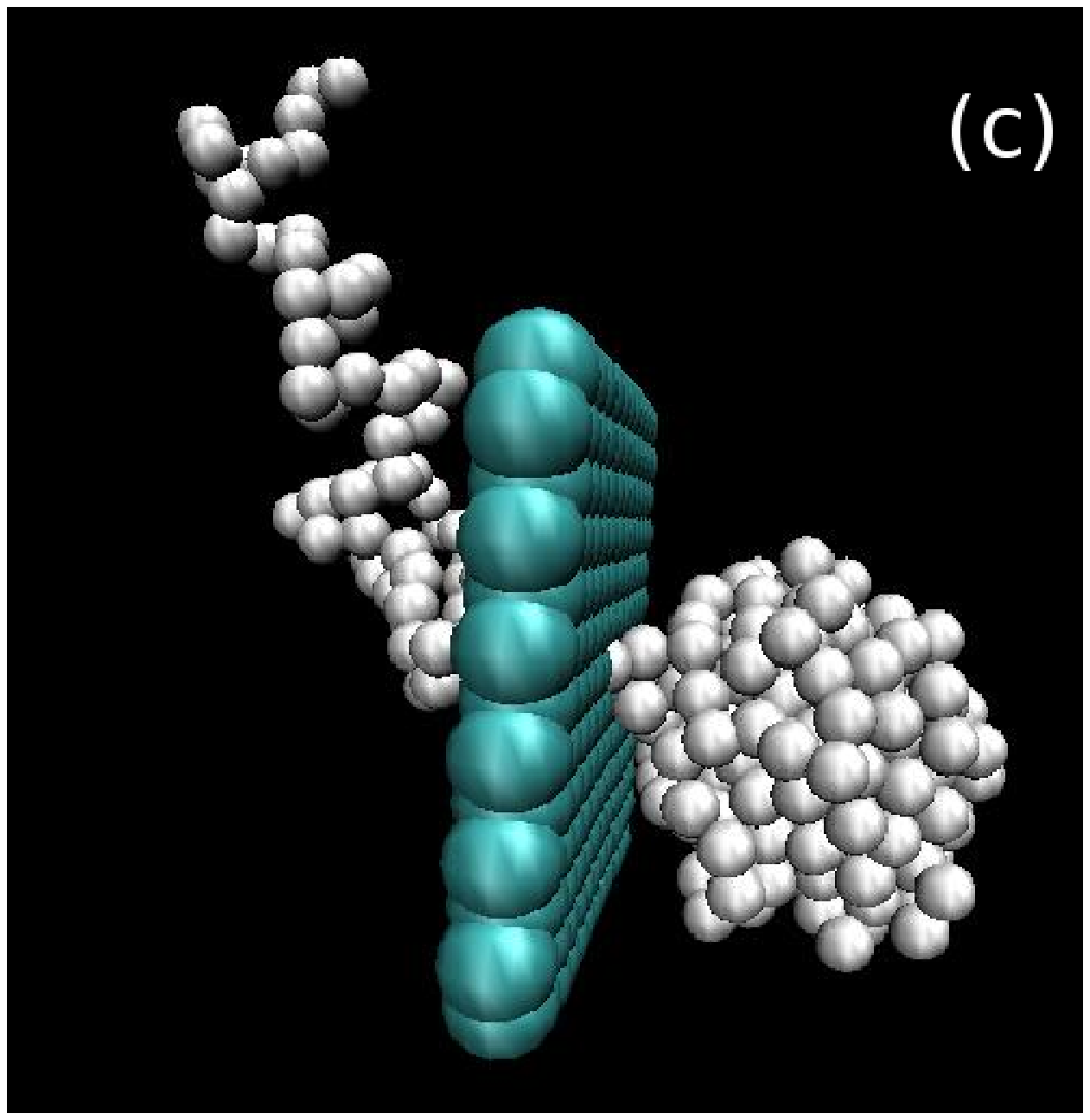}\\
\includegraphics[width=4.0truecm]{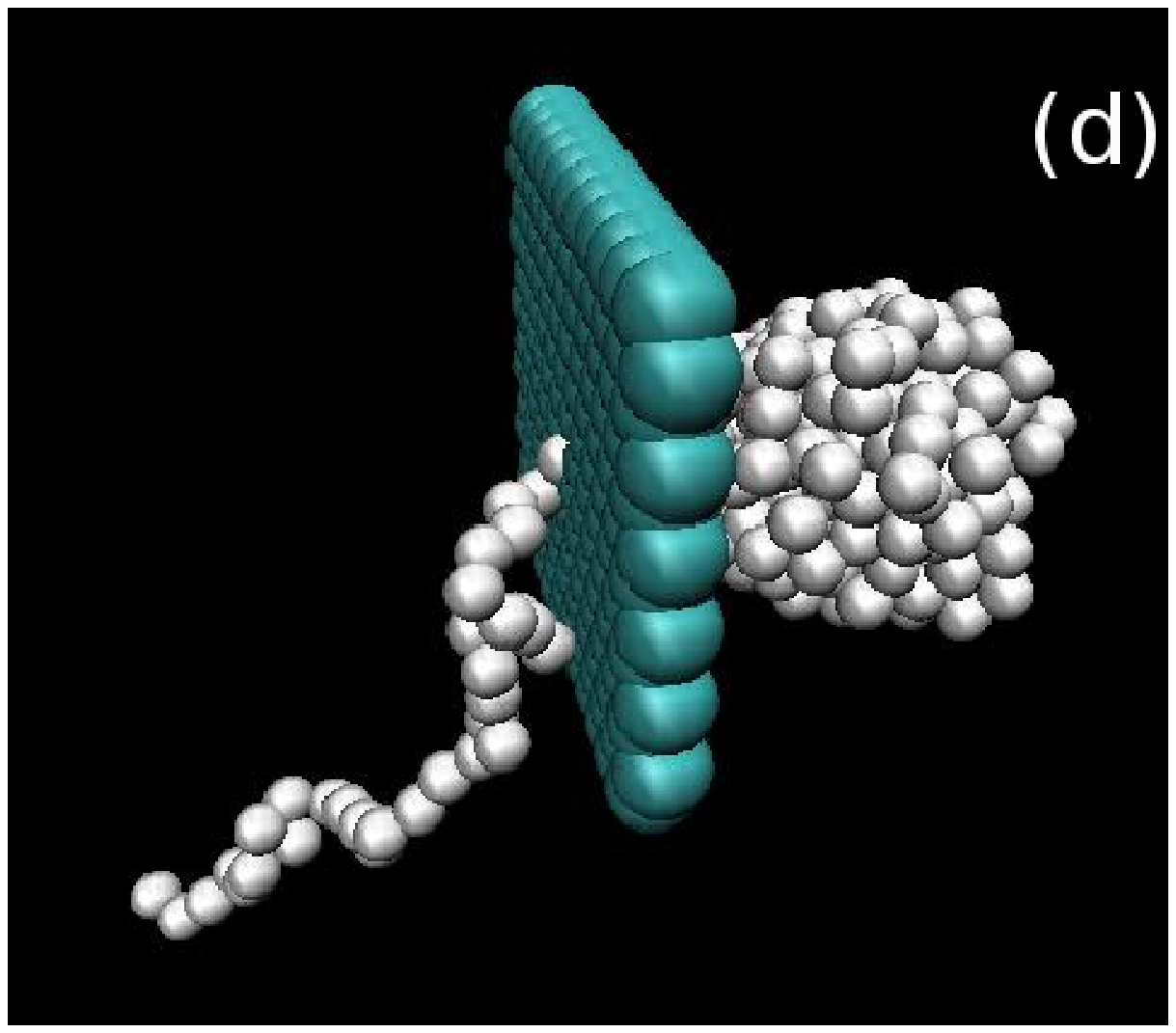}\\
\includegraphics[width=4.0truecm]{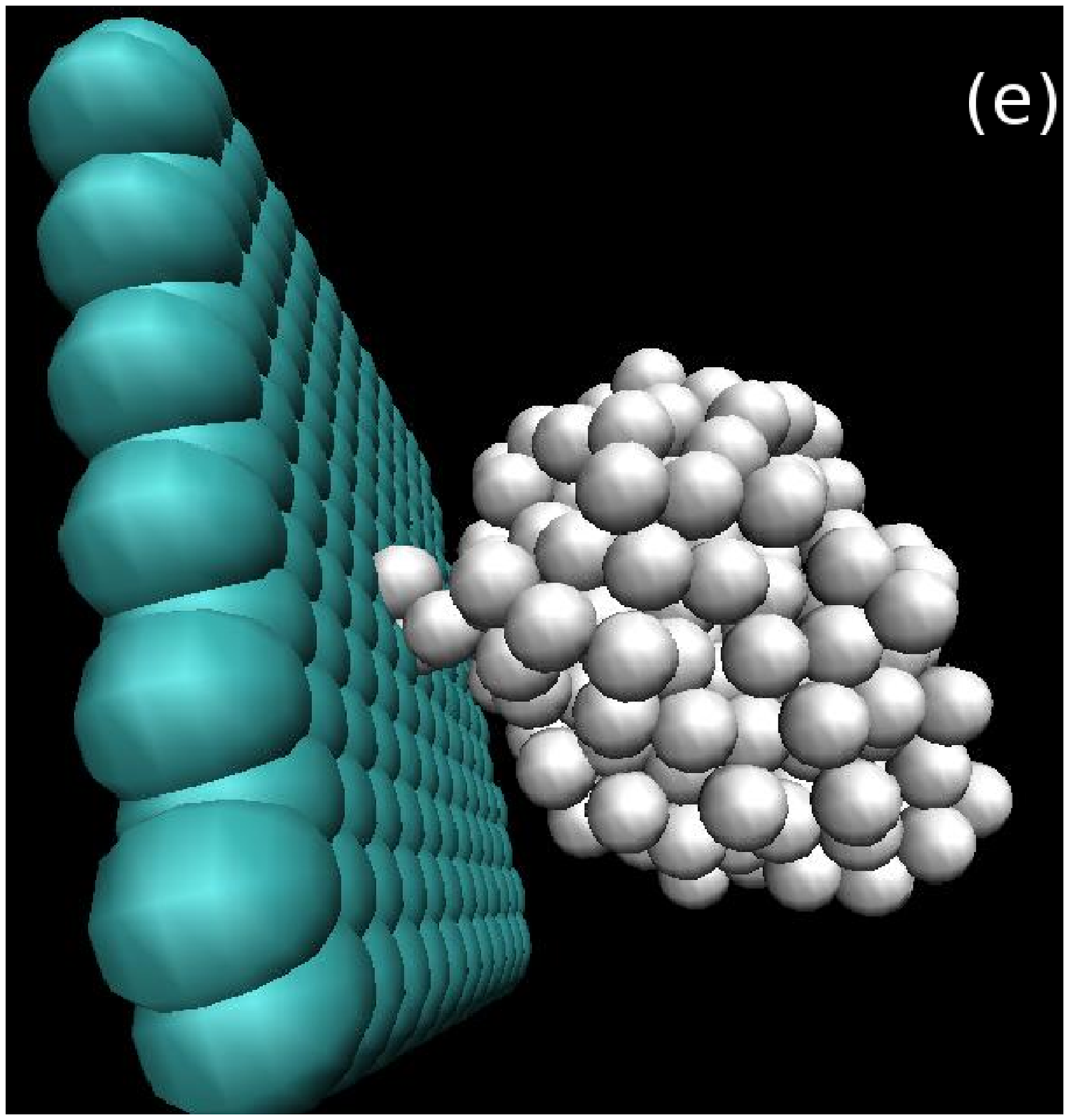}
\vskip -2.0truecm
\end{center}
\vskip 0.25truecm
\centerline{\large Fig. 2(b)}
\end{figure}
\begin{figure}[ht!]
\begin{center}
\includegraphics[width=12.0truecm]{Fig3.eps}
\label{radius_gyration}       
\end{center}
\vskip 6.0truecm
\centerline{\large Fig. 3}
\end{figure}
\begin{figure}[ht!]
\begin{center}
\includegraphics[width=11.0truecm]{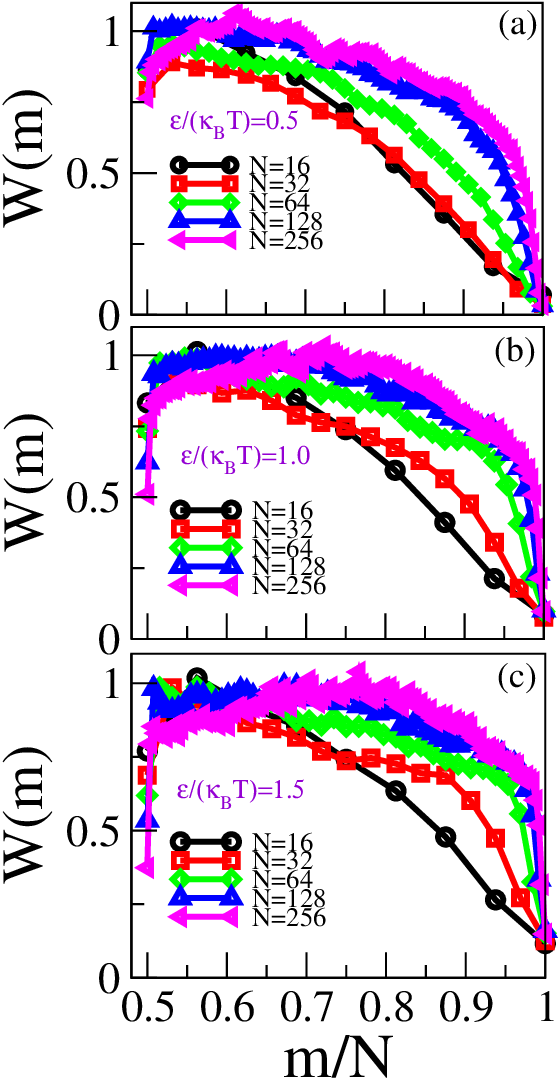}
\label{residence}
\end{center}
\vskip 0.0truecm
\centerline{\large Fig. 4}
\end{figure}
\begin{figure}[tb]                
\begin{center}
\includegraphics[width=10.0truecm]{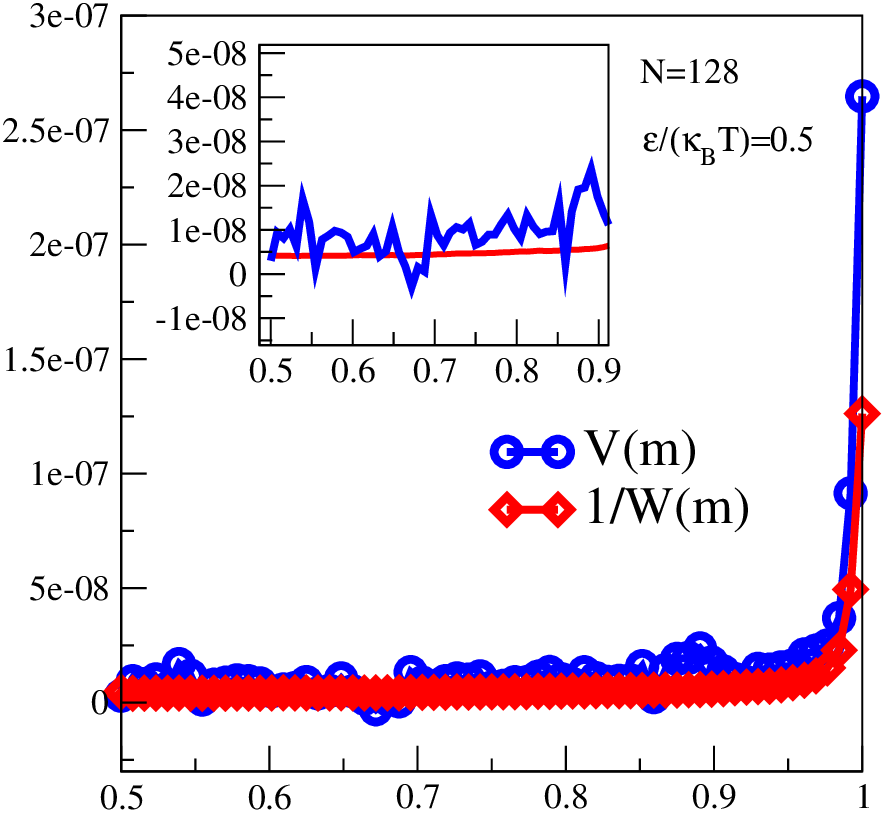}
\label{wtinv}
\end{center}
\vskip 2.5truecm
\centerline{\large Fig. 5}
\end{figure}
\begin{figure}[ht!]
\begin{center}
\includegraphics[width=11.0truecm]{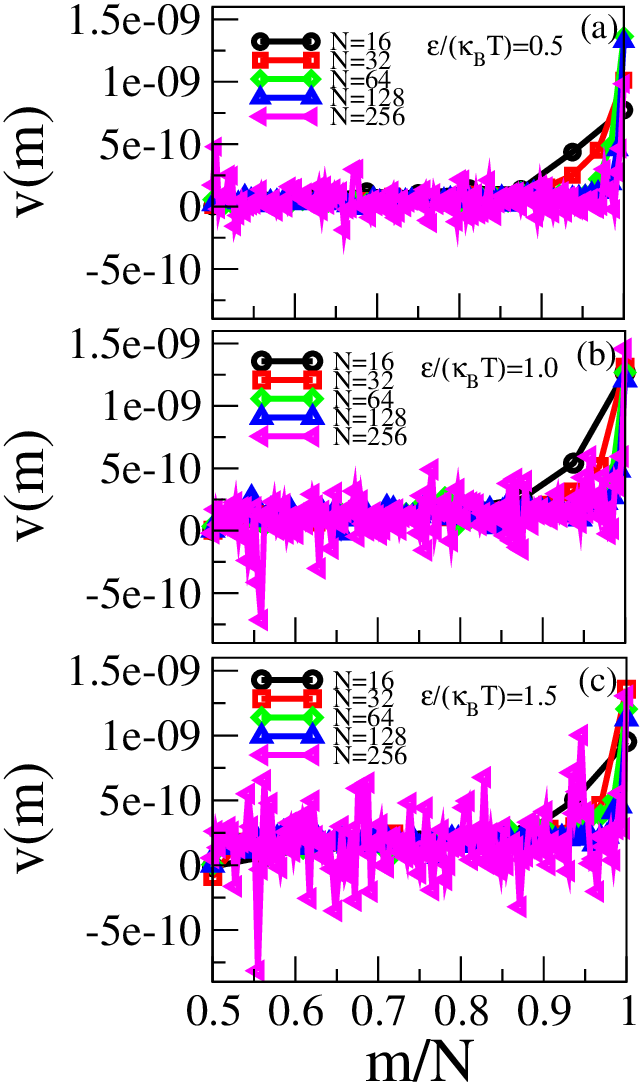}
\label{velocity}       
\end{center}
\vskip 0.0truecm
\centerline{\large Fig. 6}
\end{figure}
\begin{figure}[ht!]
\begin{center}
\includegraphics[width=12.0truecm]{Fig7.eps}
\vskip -1.0truecm
\label{force}       
\end{center}
\vskip 1.0truecm
\centerline{\large Fig. 7}
\end{figure}
\begin{figure}[tb]                
\begin{center}
\includegraphics[width=14.0truecm]{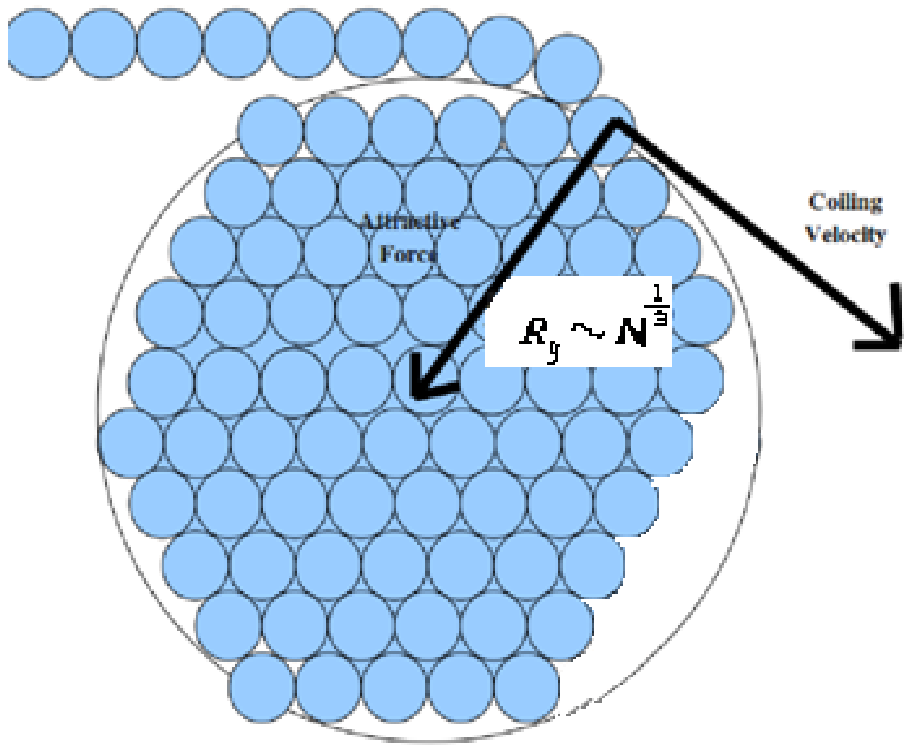}
\vskip -1.4truecm
\label{schematic}
\end{center}
\vskip 1.4truecm
\centerline{\large Fig. 8}
\end{figure}
\begin{figure}[ht!]
\begin{center}
\includegraphics[width=12.0truecm]{Fig9.eps}
\label{histo}       
\end{center}
\vskip 0.0truecm
\centerline{\large Fig. 9}
\end{figure}
\begin{figure}[ht!]
\begin{center}
\includegraphics[width=12.0truecm]{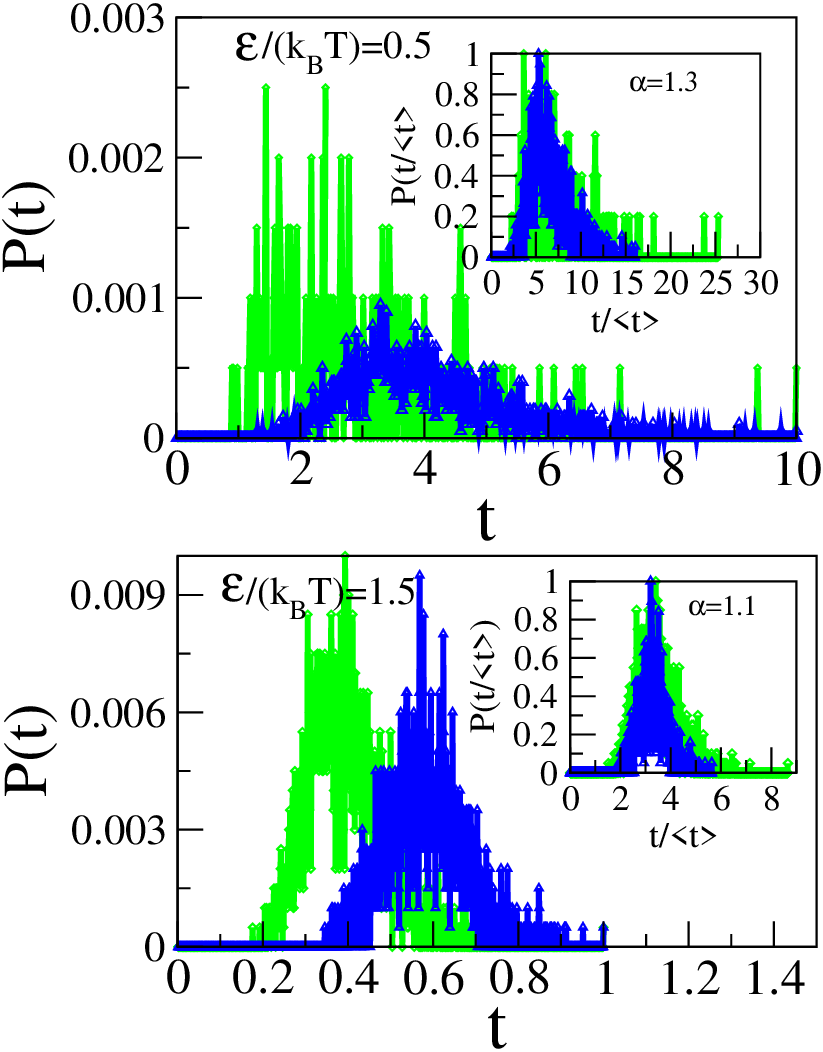}
\label{histo_diffipore}       
\end{center}
\vskip 3.0truecm
\centerline{\large Fig. 10}
\end{figure}
\begin{figure}[tb]                
\begin{center}
\includegraphics[width=9.0truecm]{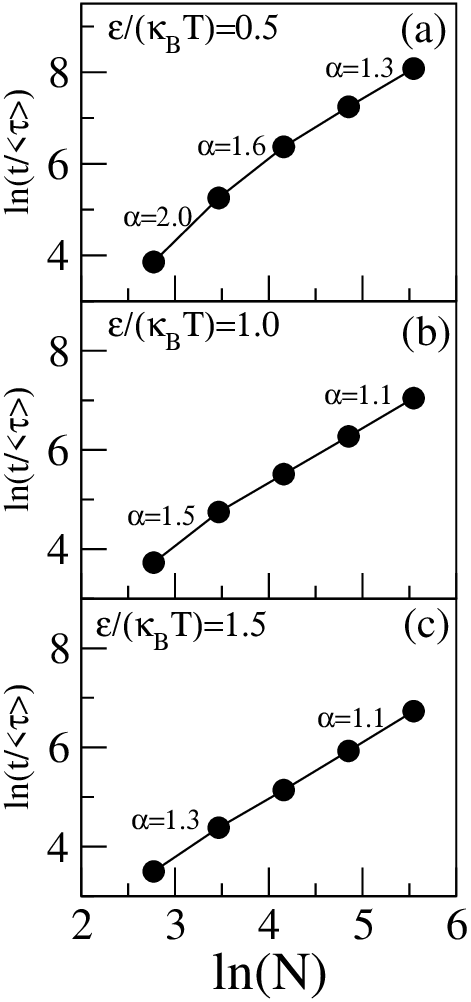}
\vskip -1.0truecm
\label{logplot}
\end{center}
\vskip 1.0truecm
\centerline{\large Fig. 11}
\end{figure}
\end{document}